\documentclass[prd,twocolumn,showpacs,amsmath,amssymb,superscriptaddress,preprintnumbers,nofootinbib,showkeys]{revtex4}

\begin{document}

\newcommand{\nablab}{{\mathop {\rule{0pt}{0pt}{\nabla}}\limits^{\bot}}\rule{0pt}{0pt}}

\title{Axionic extension of the Einstein-aether theory: \\ How does dynamic aether regulate the state of axionic dark matter?}

\author{Alexander B. Balakin}
\email{Alexander.Balakin@kpfu.ru}
\author{Amir F. Shakirzyanov}
\email{shamirf@mail.ru}
\affiliation{Department of General Relativity and
Gravitation, Institute of Physics,Kazan Federal University, Kremlevskaya street 18, Kazan, 420008,
Russia}
\date{\today}

\begin{abstract}
In the framework of axionic extension of the Einstein-aether theory we establish the model, which describes a stiff regulation of the behavior of axionic dark matter by the dynamic aether.
The aether realizes this procedure via the modified Higgs potential, designed for modeling of nonlinear self-interaction of pseudoscalar (axion) field; the modification of this potential is that its minima are not fixed, and their positions and depths depend now on the square of the covariant derivative of the aether velocity four-vector.
Exact solutions to the master equations, modified correspondingly, are obtained in the framework of homogeneous isotropic cosmological model. The effective equation of state for axionic dark matter is of the stiff type. Homogeneous perturbations of the pseudoscalar (axion) field, of the Hubble function and of the scale factor are shown to fade out with cosmological time, there are no growing modes, the model of stiff regulation is stable.

\end{abstract}
\pacs{04.20.-q, 04.40.-b, 04.40.Nr, 04.50.Kd}
\keywords{Alternative theories of gravity, Einstein-aether theory,
unit vector field, axion}
\maketitle

\section{Introduction}

The axionic extension of the Einstein-aether theory is an example of a pseudoscalar-vector-tensor model of gravity.
The  central constructive element of this theory, the unit time-like vector field $U^i$, is associated with the velocity of some cosmic substratum, indicated as the dynamic aether.
The Einstein-aether theory was created in the beginning of 21 century; the motives of its formulation, historical details and results obtained in 2000-2008 are described in the well-known Jacobson's review: the status report \cite{1}. In particular, during this first period the basic definitions and formalism of the Einstein-aether theory were established \cite{2}, the solutions describing black holes \cite{BH1,BH2,BH3}, static spherically symmetric objects, neutron stars and binary systems \cite{NS1,NS2,NS3} were obtained. In the approximation of weak field the velocities of scalar, vectorial and tensor modes in the dynamic aether were analyzed \cite{Waves}. The constraints on the quartet of coupling constants $C_1$, $C_2$, $C_3$ and$C_4$, appeared in the Einstein-aether theory are formulated (see the corresponding summary, e.g., in \cite{1}).

The next step in the development of the Einstein-aether theory was associated with its modifications, which can be formally divided into two groups: generalizations and extensions of the theory. We start to recall this story with the generalizations of the structure of the action functional of the Einstein-aether theory
\begin{equation}
S_{({\rm EA})} {=} \int \frac{d^4x \sqrt{{-}g}}{2\kappa}\left[R {+} 2\Lambda {+} \lambda(g_{ik}U^iU^k {-}1) {+} {\cal K} \right],
\label{L}
\end{equation}
where $R$ is the Ricci scalar; $\Lambda$ is the cosmological constant, and ${\cal K} = K^{ab}_{mn} \nabla_aU^m \nabla_b U^n$ is the so-called kinetic term quadratic in the covariant derivative $ \nabla_a U^m$.
The constitutive tensor $K^{ab}_{mn}$  has the form
\begin{equation}
K^{ab}_{mn}{=} C_1 g^{ab} g_{mn} {+} C_2 \delta^{a}_{m} \delta^{b}_{n}
{+} C_3 \delta^{a}_{n}\delta^{b}_{m} {+} C_4 U^{a} U^{b}g_{mn},
\label{K}
\end{equation}
and contains four coupling constants $C_1$, $C_2$, $C_3$, $C_4$ introduced phenomenologically (see \cite{1}).

The first trend in these generalizations is associated with introduction of the nonlinear term $F({\cal K})$ instead of the linear term ${\cal K}$ into the Lagrangian, ${\cal K} \to F({\cal K})$  (see, e.g., \cite{FK1,FK2}).   Clearly, the idea of such nonlinear generalization is inspired by the $f(R)$-extension of the theory of gravity, in the framework of which the Ricci scalar $R$, the basic term of the Einstein-Hilbert action functional, is replaced by the nonlinear function $R \to f(R)$.

The second type of generalization of the Einstein-aether theory is connected with the replacement of the term $\lambda(g_{ik}U^iU^k -1)$ in the Lagrangian, by the term $\lambda(g_{ik}U^iU^k -1)^2$ (see, e.g., \cite{lambda}).  In the first case $\lambda$ is considered as the Lagrange multiplier; variation with respect to $\lambda$ gives the normalization condition, and the multiplier $\lambda$ itself can be found from the field equations for $U^k$. In the second case, the quadratic term plays the role of a potential for $U_kU^k$; the quantity $\lambda$ is now an unknown constant, and the case $U^kU_k=1$ relates to the minimum of the potential; for sure, in this approach the parameter $\lambda$ enters neither the field equations for the vector field, nor the gravity field equations, thus remaining the extra parameter of the Aether model.

The generalization of the third type is predetermined by modifications in the gravitational part of the Lagrangian. For instance, there is a number of works, which exploit the modifications of the theory of gravity of the  Horava-Lifshitz type (see, e.g., \cite{H1,H2}); also, there are the modifications based on the metric-affine gravity \cite{Hehl}, on the $f(R)$, $f(G)$, $f(T)$, etc., models of gravity (see, e.g., \cite{FR}).

The fourth type of generalizations contains the theories, which involve into consideration extra-dimensions \cite{Extra1}, the ideas of supersymmetry \cite{Super1,Super2,Super3}, and the formalism of color aether \cite{color}.

When we speak about extensions of the Einstein-aether theory, we keep in mind the incorporation of new additional terms into the action (\ref{L}), which describe scalar, pseudoscalar, electromagnetic, gauge, etc., fields. The scalar extension of the Einstein-aether  theory was motivated by cosmological applications, (see, e.g., \cite{S0,S1,S2,S3,S4,S5,S6}), in particular, to precise some details of inflation, dynamics of perturbations and the large-scale structure formation. In order to describe the impact of the scalar field $\varphi$ on the unit vector field $U^k$, the authors of \cite{S0} suggested to modify the constitutive tensor $K^{ab}_{mn}$ (\ref{K}) by introduction of the functions $\beta_1(\varphi)$, $\beta_2(\varphi)$, $\beta_3(\varphi)$, $\beta_4(\varphi)$ instead of constants $C_1$, $C_2$, $C_3$, $C_4$, respectively; i.e., they used the replacement $K^{ab}_{mn} \to K^{ab}_{mn}(\varphi)$. The backreaction  of the aether on the evolution of the scalar field is described in \cite{S1} by introduction of modified scalar field potential $V(\varphi,\Theta)$ depending on the expansion scalar $\Theta \equiv \nabla_k U^k$; this idea was developed, and, e.g., in \cite{S6} the modified potential $V(\varphi,\Theta, \sigma^2)$ appeared, where $\sigma^2 \equiv \sigma_{mn}\sigma^{mn}$ is the square of the shear tensor.

The electromagnetic extension of the Einstein-aether theory, which takes into account terms up to the fourth order in derivatives advocated by the Effective field theory \cite{EFT1,EFT4}, was established in \cite{EM1} and applied to the problem of birefringence in \cite{EM2}. The $SU(N)$ generalization of the Einstein-Maxwell-aether theory and description of the aether coupling to the Yang-Mills fields were presented in \cite{color}.

Pseudoscalar (axionic) extension of the Einstein-aether theory was presented in the work \cite{A1}, in which all the possible (irreducible) terms up to second order in derivatives were included into the Lagrangian and analyzed. The extension of this type was motivated by the interest to the cosmological models with the pseudoscalar (axion) field $\phi$ associated with the axionic dark matter (see, e.g., \cite{PQ,Weinberg,Wilczek,Ni77,ADM1,ADM2,ADM3,ADM4,ADM41,ADM5,ADM6,ADM7,ADM8} for review, references, historical and mathematical details).

In many respects the model with pseudoscalar field is similar to the model with scalar field as the third ingredient, however, there is a number of dissimilarities \cite{EM1}. The main difference between the models with scalar and pseudoscalar (axion) fields is that in the second case, respectively, only even functions of $\phi$ can be introduced into the Lagrangian of the model. In particular, this means that the potential of the axion field has to be written in the form $V(\phi^2)$; as a consequence, one can not use the linear function $\alpha \phi$, the exponential potentials $e^{\lambda \phi}$ (but, of course, one can consider $\cos{\mu \phi}$, $\cosh{\lambda \phi}$, $\phi^{2m}$, etc...). Another difference appears, when we add terms with Levi-Civita pseudo-tensor $\epsilon^{ikmn}$ into the Lagrangian of the model \cite{A1}. In this case, for instance, the convolution of the gradient four-vector of the pseudoscalar field $\nabla_i \phi$  with the angular velocity pseudo-four vector ($\omega^i = \frac12 \epsilon^{ikmn}U_k \nabla_m U_n$) gives the true scalar  $\omega^i \nabla_i \phi$. When the electromagnetic field is incorporated into the models with pseudoscalar (axion) field, a new invariant appears in the total Lagrangian, $\phi F^{*}_{mn}F^{mn}$ with the dual Maxwell tensor $F^{*}_{mn} =\frac12 \epsilon_{mnpq}F^{pq}$ \cite{Ni77}. Similar situation appears, when one replaces the $U(1)$ symmetric Maxwell tensor $F_{mn}$  with the $SU(N)$ symmetric Yang-Mills field ${\bf F}_{mn}$. In other words, the pseudoscalar extension of the Einstein-aether theory appeared for description of the aether coupling to the axionic dark matter, can not be reduced to the scalar extension of this theory.

The next natural step in the development of the Einstein-aether theory was the combination of the electromagnetic and pseudoscalar modifications indicated as axionic extension of the Einstein-Maxwell-aether theory \cite{EMA1,EMA2,EMA3}; in particular, such an extension gave the possibility to describe  dynamo-optical anomalies in the electromagnetic response of the axionically active systems \cite{EMA1}.

It is well-known that the presence of the unit dynamic time-like vector-field $U^k$ realizes the idea of a preferred frame of reference (see, e.g., \cite{CW,N1,N2,N3}), and violates the Lorentz invariance of the model (see, e.g., \cite{LV1,LV2} for discussions about the consequences of the broken Lorentz invariance). During last three years the discussion on this topic was revived due to the interest to the so-called universal horizons and thermodynamics of the Einstein-aether black holes (see, e.g., \cite{UH1,UH2,UH3}).

In 2017 the outstanding events, encoded as GW170817 and GRB 170817A \cite{170817}, changed cardinally the status of constraints on the Einstein-aether theory and its modifications. Based on the observation of the binary neutron star merger, it was established that the ratio of the velocities of the gravitational and electromagnetic waves differs from one by the quantity about $10^{-15}$ (to be more precise, $1-3 \times 10^{-15}< \frac{v_{\rm gw}}{c}<1+ 7 \times 10^{-16} $). Since the square of the velocity of the tensorial spin-2 aether mode was calculated in \cite{Waves} to be $S^2_{(2)}= \frac{1}{1-(C_1+C_3)}$, one can estimate the sum of the parameters $C_1+C_3$ as follows: $-6 \times 10^{-15}<C_1+C_3< 1.4 \times 10^{-15}$. In other words, we can consider that   $S_{(2)}=1$ with very high precision, and assume that $C_3 \simeq -C_1$. Correspondingly, the square of the velocity of the vectorial spin-1 aether mode is $S^2_{(1)}= \frac{C_1}{C_1+C_4}$; since this square is nonnegative, we can see that $\frac{C_4}{C_1}>-1$, if $C_1 \neq 0$, or $C_1=0$, if $S_{(1)}=0$. Similarly, the square of the velocity of the scalar spin-0 mode can be rewritten as $S^2_{(0)}= \left[\frac{2}{C_1+C_4}-1\right] / \left[3+ \frac{2}{C_2} \right] \geq 0$. In other words, one has now one new constraint, which is experimentally proved, and two its consequences.

In this paper we are working in the context of axionic extension of the Einstein-aether theory \cite{A1}.  In order to describe the impact of the aether on the pseudoscalar (axion) field we use the idea, which was proposed in \cite{S1,S6} for the scalar extension of the Einstein-aether theory. Let us recall, that in \cite{S1} the potential of the scalar field $\varphi$ was chosen as a function of the expansion scalar $\Theta$, i.e., $V(\varphi, \Theta)$; in the framework of spatially isotropic cosmological model the only $\Theta$ is the nonvanishing scalar containing the covariant derivative of the velocity four-vector $U^k$. The authors of \cite{S6} have used the potential $V(\varphi, \Theta, \sigma^2)$ supplemented by the square of the shear tensor $\sigma^2$; this supplementary argument is not vanishing, first, in the anisotropic cosmological models of the Bianchi-I type, second, in the models with the pp-wave symmetry (see, e.g., \cite{EM2}).
The nonvanishing vorticity tensor $\omega^{ik}$ associated with the aether unit vector field appears in the context of rotating cosmological model of the G\"odel type \cite{S7}; this motif could be interesting if there is an idea to include the argument $\omega^2 \equiv \omega_{ik} \omega^{ik}$ into the potential of the scalar field $\varphi$ or of the pseudoscalar (axion) field. The acceleration four-vector $d^k \equiv U^n \nabla_n U^k$ is nonvanishing in the static spherically symmetric models; it would be interesting to study the aether vector field influence on the axionic dark matter, by using the new argument $d^2 = U^m \nabla_m U_k U^n \nabla_n U^k$ in the potential of the pseudoscalar field along the line proposed in \cite{Dyon}.
To conclude, we would like to stress that there exists an interest and serious motivation to consider the potential of the pseudoscalar (axion) field of the following type: $V(\phi^2, \Theta^2, \sigma^2, \omega^2, d^2)$
There are two evident ways to simplify the construction of the potential of this type by agglomerating the quantities $\Theta^2, \sigma^2, \omega^2, d^2$. The first way is to use the potential $V(\phi^2, {\cal K})$, where ${\cal K}$ is the already introduced kinetic term containing the unknown Jacobson's coupling constants (\ref{K}). The second way is to introduce only one (cumulative) dynamic scalar $\Omega^2 \equiv \nabla_m U_n \nabla^m U^n$, which is, in fact, the sum $\Omega^2= d^2+ \sigma^2 + \omega^2 + \frac13 \Theta^2$, and does not contain any extra parameters, which could divide the contribution of acceleration, shear, vorticity and expansion of the aether velocity flow. Some extra arguments and motivation for such choice can be found in the work \cite{BD}.

We assume the modified potential of the pseudoscalar field  $V(\phi^2, \Omega^2)$ to be of the Higgs type, $V = \frac12 \gamma \left(\phi^2{-}\Phi^2_{*} \right)^2$.  The dynamic scalar $\Omega^2$ is assumed to be incorporated into the the quantities $\pm \Phi_{*}$, which correspond to the pair of minima of this potential and thus describe two basic states of the axion field, i.e., $\Phi_{*}=\Phi_{*}(\Omega^2)$. This potential appears in the class of models, which can be indicated as models with $\phi^4$-type self-interaction; this potential is widely used in cosmology for  description of the axionic dark matter (see, e.g., \cite{ADM41,Sikivie2}).
Clearly, in the context of isotropic cosmology, the new variable in the potential of the axion field is reduced to the expansion scalar $\Theta$ only, $\Omega^2 \to \frac13 \Theta^2$, thus recovering the idea presented in \cite{S1}.

We have shown in \cite{EMA3} that the influence of the axionic dark matter on the dynamic aether can switch on or switch off the pp-wave modes. Now we focus on the inverse effect, when the dynamic aether guides the behavior of the axionic dark matter via the dynamic scalar $\Omega^2$ incorporated into the potential of the pseudoscalar field, $V(\phi^2, \Omega^2)$.

The paper is organized as follows. In Section II, we recall the basic elements of formalism, and derive master equations for pseudoscalar, unit vector and gravitational fields. In Section III we consider cosmological applications of the axionic extension of the Einstein-aether theory, and obtain new exact solutions for the spatially isotropic homogeneous Universe with and without cosmological constant. Section IV contains discussion and conclusions.

\section{Formalism of axionic extension of the Einstein-aether theory}

\subsection{The action functional}

We consider the action functional of the axionic extension of the Einstein-aether theory in the following form:
$$
S_{({\rm EAA)}} =  \int d^4 x \sqrt{{-}g} \ \left\{ \frac{1}{2\kappa}\left[R{+}2\Lambda {+} \lambda (g_{mn}U^m
U^n {-}1 ){+} \right.\right.
$$
\begin{equation}
 \left. \left.   K^{ab}_{mn} \nabla_a U^m \nabla_b U^n \right] {+}
\frac{1}{2}\Psi^2_0 \left[V(\phi^2,\Omega^2) {-} g^{mn}\nabla_m \phi \nabla_n \phi \right] \right\}
.
\label{total2}
\end{equation}
In addition to quantities, described above, this action functional includes the dimensionless quantity $\phi$ denoting the pseudoscalar (axion) field. The parameter $\Psi_0$ is reciprocal
to the constant of the axion-photon coupling $g_{({\rm A} \gamma \gamma)} {=}\frac{1}{\Psi_0}$; the constraint for the constant $g_{({\rm A} \gamma \gamma)}$ is  $g_{({\rm A} \gamma \gamma)}< 1.47 \times 10^{-10} {\rm GeV}^{-1}$ (see \cite{CAST14}).

The second argument of the potential of the pseudoscalar field, $V(\phi^2, \Omega^2)$, the scalar $\Omega^2 \equiv \nabla_{m} U_n \nabla^{m} U^n$, can be represented using the standard
decomposition of the covariant derivative of the velocity field $U^i$
\begin{equation}
\nabla_i U_k = U_i DU_k + \sigma_{ik} + \omega_{ik} +
\frac{1}{3} \Delta_{ik} \Theta \,. \label{act3}
\end{equation}
The acceleration four-vector $DU^i$, the symmetric shear tensor $\sigma_{mn}$,  the skew-symmetric vorticity tensor $\omega_{mn}$, and the expansion scalar $\Theta$ are the irreducible elements
of this decomposition defined as follows:
$$
DU_k \equiv  U^m \nabla_m U_k \,, \quad \sigma_{ik}
\equiv \frac{1}{2}\left(\nablab_i U_k {+}
\nablab_k U_i \right) {-} \frac{1}{3}\Delta_{ik} \Theta  \,,
$$
$$
\omega_{ik} \equiv \frac{1}{2} \left(\nablab_i U_k {-} \nablab_k U_i \right) \,, \quad \Theta \equiv \nabla_m U^m
\,,
$$
\begin{equation}
D \equiv U^i \nabla_i \,, \quad \Delta^i_k = \delta^i_k - U^iU_k \,, \quad \nablab_i \equiv \Delta_i^k \nabla_k \,. \label{act4}
\end{equation}
In these terms the scalar $\Omega^2$ takes the form
\begin{equation}
\Omega^2 = DU_m DU^m + \sigma_{mn} \sigma^{mn} +  \omega_{mn} \omega^{mn} + \frac13 \Theta^2 \,, \label{dec33}
\end{equation}
and the kinetic term ${\cal K}$ is
$$
{\cal K} \equiv K^{abmn}(\nabla_a U_m) (\nabla_b U_n) =
$$
$$
=(C_1 {+} C_4)DU_k DU^k {+}
(C_1 {+} C_3)\sigma_{ik} \sigma^{ik} {+}
$$
\begin{equation}
+ (C_1 {-} C_3)\omega_{ik}
\omega^{ik} {+} \frac13 \left(C_1 {+} 3C_2 {+}C_3 \right) \Theta^2
\,. \label{act5n}
\end{equation}
Clearly, the scalar (\ref{act5n}) coincides with $\Omega^2$, when $C_1=1$ and $C_2=C_3=C_4=0$.

\subsection{Master equations for the pseudoscalar, unit vector, and gravitational fields}

\subsubsection{Master equation for the axion field}

Variation of the action functional $S_{({\rm EAA)}}$ (\ref{total2}) with respect to pseudoscalar field $\phi$ gives the master equation in the standard form
\begin{equation}
g^{mn} \nabla_m \nabla_n \phi = -  \phi  \frac{\partial V}{\partial \phi^2} \,.
\label{ax10}
\end{equation}
We assume the potential of the axion field is of the Higgs type
\begin{equation}
V(\phi^2,\Omega^2 )= \frac12 \gamma \left[\phi^2 - \Phi^2_{*}(\Omega^2) \right]^2
\,.
\label{P1}
\end{equation}
This potential has the maximum located at $\phi = 0$), and two symmetric minima located at $\phi = \pm \Phi_{*}$. The positions of these minima, as well as, the height of the barrier between the maximum and minimum, $V_0 = \frac12 \gamma \Phi^4_{*}(\Omega^2)$, are not fixed and depend on the value of the function $\Phi_{*}(\Omega^2)$. When $\phi = \Phi_{*} {+} \psi$, $|\psi|<<1$, i.e., the deviation of axion field from the minimal value is small, the potential reduces to $2\gamma \Phi^2_{*} \phi^2$, so that the quantity $2\gamma \Phi^2_{*}$ plays the role of square of an effective mass, $m^2_{(\rm A)}=2\gamma \Phi^2_{*}$.  We would like to repeat that the quantity $\Phi_{*}(\Omega^2)$ is not now the constant, it depends, generally speaking, on time and spatial coordinates through the covariant derivative of the aether vector field. Thus, we deal with the case, when the dynamic aether controls the behavior of the axionic dark matter. To be more precise, the aether prescribes what is the basic state of the axion field; alternatively, one can say that the aether regulates the effective mass of the axion field.
Thus the equation
\begin{equation}
\nabla_m \nabla^m \phi = -  \gamma \phi\left[\phi^2 - \Phi^2_{*}(\Omega^2) \right]
\label{ax1}
\end{equation}
is the master equation for the pseudoscalar field.

\subsubsection{Equations for the unit dynamic vector field}

The aether dynamic equations can be found by variation of the action (\ref{total2}) with
respect to the Lagrange multiplier $\lambda$ and to the unit vector field $U^i$.
The variation of the action (\ref{total2}) with respect to
$\lambda$ yields the equation
\begin{equation}
g_{mn}U^m U^n = 1 \,,
\label{21}
\end{equation}
which is known to be the normalization condition of the time-like
vector field $U^k$.
Then, variation of the functional (\ref{total2}) with respect to
$U^i$ yields the master equation in the standard form:
\begin{equation}
\nabla_a {\cal J}^{aj}
 = \lambda \ U^j  + I^j \,,
\label{0A1}
\end{equation}
where $I^j$ as in \cite{2} is of the form
\begin{equation}
I^j =  C_4 (DU_m)(\nabla^j U^m) \,.
\label{J7}
\end{equation}
The essential difference from \cite{2} appears in the term ${\cal J}^{aj}$, which is defined now as follows:
$$
{\cal J}^{aj} = \tilde {K}^{abjn} (\nabla_b U_n)  \,,
$$
\begin{equation}
\tilde{K}^{abjn} = K^{abjn} + \kappa \Psi^2_0 \ g^{ab}g^{jn} \frac{\partial V}{\partial \Omega^2} \,,
\label{J2}
\end{equation}
where the tensor $K^{abjn}$ is the rewritten tensor (\ref{K}).
In fact, one can obtain the new constitutive tensor $\tilde{K}^{abmn}$ from the Jacobson's one (\ref{K}), if we replace the constant $C_1$ by the function $h_1(\phi^2, \Omega^2)$ defined as
\begin{equation}
h_1(\phi^2, \Omega^2) = C_1 -2 \kappa \gamma \Psi^2_0 \left[\phi^2 - \Phi^2_{*}(\Omega^2) \right] \Phi_{*} \frac{d \Phi_{*}}{d \Omega^2} \,.
\label{J29}
\end{equation}
The Lagrange multiplier $\lambda$ can be obtained standardly as
\begin{equation}
\lambda =  U_j \left[\nabla_a {\cal J}^{aj}- I^j \right]  \,,  \label{0A309}
\end{equation}
it also depends on the guiding function $\Omega^2$. Let us mention that the presence of the term $h_1(\phi^2, \Omega^2)$ in the functions ${\cal J}^{aj}$ and $\lambda$ signals that there exists, in general case, the backreaction of the axion field, controlled by the aether, on the unit vector field evolution.

\subsubsection{Equations for the gravitational field}

The variation of the action (\ref{total2}) with respect to the metric
$g^{ik}$ yields the gravitational field equations, which can be presented in the following form
\begin{equation}
R_{ik} - \frac{1}{2} R \ g_{ik}
=  \Lambda g_{ik}  + T_{ik} +
\kappa T^{({\rm A})}_{ik} \,. \label{0Ein1}
\end{equation}
$$
T_{ik} =
\frac12 g_{ik} \ K^{abmn} \nabla_a U_m \nabla_b U_n{+} U_iU_k U_j \nabla_a {\cal J}^{aj} {+}
$$
$$
{+}\nabla^m \left[U_{(i}{\cal J}_{k)m} {-}
{\cal J}_{m(i}U_{k)} {-}
{\cal J}_{(ik)} U_m\right]+
$$
$$
+h_1\left[(\nabla_mU_i)(\nabla^m U_k) {-}
(\nabla_i U_m )(\nabla_k U^m) \right] {+}
$$
\begin{equation}
{+}C_4 \left[D U_i D U_k - U_iU_k DU_m DU^m \right] \,.
\label{5Ein1}
\end{equation}
As usual, the symbol $p_{(i} q_{k)}$
denotes symmetrization.
The quantity $T^{({\rm A})}_{ik}$ written as
\begin{equation}
T^{({\rm A})}_{ik} = \Psi^2_0 \left[\nabla_i \phi \nabla_k \phi
+\frac12 g_{ik}\left(V {-} \nabla_n \phi \nabla^n \phi \right) \right]
\label{qq1}
\end{equation}
describes the extended stress-energy tensor of the pseudoscalar field.
Compatibility conditions for the set of equations (\ref{0Ein1})
\begin{equation}
\nabla^k\left[T_{ik} +
\kappa T^{({\rm A})}_{ik} \right] = 0
\,, \label{compa1}
\end{equation}
are satisfied automatically on the solutions to the master equations (\ref{ax1})-(\ref{0A309}).

\section{The application: Spatially isotropic and homogeneous cosmological model with dynamic aether and axionic dark matter}

\subsection{Reduced master equations}

In this Section we consider the master equations for the pseudoscalar, vector and gravitational field for the symmetry associated with spatially isotropic,
 homogeneous cosmological model of the Friedmann type. We assume the metric to be of the form
\begin{equation}
ds^2 = dt^2 - a^2(t)[dx^2+dy^2+dz^2] \,,
\label{App1}
\end{equation}
with the scale factor $a(t)$ and the Hubble function $H(t){=}\frac{\dot{a}}{a}$. The dot denotes the derivative with respect to cosmological time $t$; we use the units with $c=1$.
We use the ansatz that the pseudoscalar and unit dynamic vector fields, $\phi$ and $U^i$, inherit the chosen symmetry. Mathematically, this requirement means that the pseudoscalar and vector fields are the functions of the cosmological time only, $\phi(t)$ and $U^i(t)$, and the velocity four-vector has to be of the form $U^i = \delta^i_0$, thus providing the absence of preferred spatial directions in the isotropic spacetime.

The covariant derivative $\nabla_i U_k$ in this case is characterized by vanishing acceleration four-vector, shear and vorticity tensors:
\begin{equation}
DU^i = 0 \,, \quad \sigma_{mn}=0\,, \quad \omega_{mn}=0 \,.
\label{App2}
\end{equation}
Only the expansion scalar is nonvanishing, and we can write
\begin{equation}
\Theta = 3H(t) \,, \quad \nabla_i U_k =  \Delta_{ik} \ H(t) \,, \quad \Omega^2 = 3H^2 \,.
\label{App3}
\end{equation}
Our first task is to prove that for our ansatz the evolutionary equations for the unit vector field  (\ref{0A1}) are satisfied identically; then we will consider the reduced equations for pseudoscalar and gravitational fields, and obtain exact solutions to these equations.

\subsubsection{Reduced equations for the unit vector field}

Using the ansatz about the velocity four-vector, $U^i{=}\delta^i_0$, we can calculate explicitly all the necessary quantities. First, we see that
the four-vector $I^j$ vanishes. Second, we obtain that
\begin{equation}
{\cal J}^{aj} {=} H \left[\Delta^{aj}\left(h_1{+}3C_2{+}C_3 \right) {+} 3C_2 U^a U^j \right]  \,,
\label{App3R}
\end{equation}
and the divergence four-vector $\nabla_a {\cal J}^{aj}$ is parallel to the velocity four-vector
\begin{equation}
\nabla_a {\cal J}^{aj} = 3U^j \left[C_2\dot{H} - H^2 \left(h_1+C_3 \right) \right] \,.
\label{App8}
\end{equation}
Thus, three of four basic evolutionary equations (\ref{0A1}) for the unit vector field are satisfied identically, and the last one defines the Lagrange multiplier (see (\ref{0A309})):
\begin{equation}
\lambda(t) =
-3H^2 \left(h_1+C_3 \right) + 3C_2 \dot{H}
\,.
\label{App11}
\end{equation}

\subsubsection{Reduced equation for the pseudoscalar (axion) field}

The reduced evolutionary equation for the axion field
\begin{equation}
\left[\ddot{\phi} + 3H \dot{\phi} \right] +  \gamma \phi\left[\phi^2 - \Phi^2_{*} \right] = 0
\label{App125}
\end{equation}
is, in general case, the non-linear one.

\subsubsection{Reduced equations for the gravitational field}

The stress-energy tensor describing the contribution of the unit dynamic vector field is
$$
T_{ik} = \Delta_{ik}\left\{\frac13 \kappa \gamma \Psi^2_0 \frac{d}{dt}\left[\left(\phi^2{-}\Phi^2_{*} \right) \frac{\Phi_{*}}{H}\frac{d\Phi_{*}}{dH} \right] + \right.
$$
$$
\left. + \frac32 H^2 \left(C_1{+}3C_2{+}C_3 \right) - \left(\dot{H} +3H^2 \right) \left(h_1{+}3C_2{+}C_3 \right)\right\} +
$$
$$
 + U_i U_k \left[ \gamma \kappa \Psi^2_0 H \Phi_{*}\left(\phi^2{-}\Phi^2_{*} \right)\frac{d\Phi_{*}}{dH} - \right.
$$
\begin{equation}
\left. - \frac32 H^2 \left(C_1{+}3C_2{+}C_3 \right) \right] \,.
\label{App14}
\end{equation}
The stress-energy tensor of the pseudoscalar field reads
\begin{equation}
T^{(\rm A)}_{ik} = \frac12 \Psi^2_0 \left[U_iU_k \left(V+{\dot{\phi}}^2\right) + \Delta_{ik} \left(V-{\dot{\phi}}^2\right)\right]\,.
\label{App16}
\end{equation}
As usual, the scalars $W_{(\rm A)}$ and ${\cal P}_{(\rm A)}$, given by
\begin{equation}
W_{(\rm A)} =   \frac12 \Psi^2_0 \left(V{+}{\dot{\phi}}^2\right) \,,  \quad {\cal P}_{(\rm A)} = \frac12 \Psi^2_0 \left({\dot{\phi}}^2 {-} V \right) \,,
\label{TA1}
\end{equation}
describe, respectively, the energy density and pressure of the axionic dark matter \cite{Odin}.

\subsubsection{The key equation for the gravity field}

As usual, only one of two non-trivial equations for the gravity field is independent for the symmetry associated with the Friedmann-type model; the second equation is the differential consequence, since the compatibility conditions are satisfied identically for the solutions to the axion field equation. This key equation can be written in the following form:
$$
\frac{1}{\kappa \Psi^2_0 } \left\{3H^2 \left[1+ \frac12 \left(C_1{+}3C_2{+}C_3 \right) \right] {-} \Lambda \right\} =
$$
\begin{equation}
= \frac12 {\dot{\phi}}^2 {+} \frac14 \gamma \left(\phi^2{-}\Phi^2_{*}\right)^2 {+}  \gamma H \Phi_{*}\left(\phi^2{-}\Phi^2_{*} \right)\frac{d\Phi_{*}}{dH}\,.
\label{App186}
\end{equation}

\subsection{Exact solution, describing the basic state of axion field in the case $\Lambda \neq 0$}

We indicate the state of the pseudoscalar (axion) field as the basic one, when $\phi=\pm \Phi_{*}$; in this sense the evolution of the function $\Phi_{*}(t)$, guided by non-uniformly moving aether, describes the evolution of the axionic dark matter. For definiteness, below we assume that $\phi= {+} \Phi_{*}$.  Now the model is reduced to the following pair of evolutionary equations:
\begin{equation}
\ddot{\Phi}_{*} + 3H \dot{\Phi}_{*} = 0 \,,
\label{HH1}
\end{equation}
\begin{equation}
H^2  = H^2_{\infty} +  \frac{\kappa \Psi^2_0}{6\Gamma}  \ {\dot{\Phi}_{*}}^2 \,.
\label{HH2}
\end{equation}
Here we introduced two convenient parameters: the first one is $\Gamma$, the parameter containing the Jacobson's constants only:
\begin{equation}
\Gamma \equiv 1+ \frac12 \left(C_1{+}3C_2{+}C_3 \right) \,.
\label{HH3}
\end{equation}
The second parameter, $H_{\infty}$, given by
\begin{equation}
H_{\infty} \equiv \sqrt{\frac{\Lambda}{3\Gamma}} \,,
\label{HH4}
\end{equation}
plays the role of asymptotic value of the Hubble function, if the $\dot{\Phi}_{*}(t \to \infty) \to 0$.

\subsubsection{Searching for geometrical characteristics of the model}

If we extract $\dot{\Phi}_{*}$ from (\ref{HH2}) and put it into (\ref{HH1}), we obtain the equation for the Hubble function
\begin{equation}
\dot{H} + 3\left(H^2-H^2_{\infty} \right)=0 \,,
\label{HH5}
\end{equation}
the solution of which is
\begin{equation}
	H(t)=H_{\infty} \ cth{[3H_{\infty}(t-t_*)]}\,.
\label{HH6}
\end{equation}
Using (\ref{HH6}) we obtain the scale factor to have the form
\begin{equation}
	a(t)=a_* \left\{ sh\left[3H_{\infty}(t-t_*)\right]\right\}^{\frac13}\,.
\label{HH7}
\end{equation}
The acceleration parameter is
\begin{equation}
-q(t) \equiv \frac{\ddot{a}}{a H^2} = 1+\frac{\dot{H}}{H^2} = 1 - \frac{3}{ch^2[3H_{\infty}(t-t_{*})]}
\label{HH8}
\end{equation}
The new parameters $a_{*}$  and $t_{*}$ are formal integration constants, which will be identified in the next subsubsection.

\subsubsection{Searching for the solution for the pseudoscalar field}

The quantity  $\dot{\Phi}_{*}(t)$ as the function of time at $t>t_{*}$ can be extracted from (\ref{HH2}) and (\ref{HH6}):
\begin{equation}
	\dot{\Phi}_*(t)= \pm \sqrt{\frac{2\Lambda}{\kappa \Psi^2_0}} \left\{sh \left[3H_{\infty}(t-t_{*})\right]\right\}^{-1}\,,
\label{HH21}
\end{equation}
clearly, it vanishes asymptotically, $\dot{\Phi}_*(\infty)=0$, for both signs in front of square root.
Respectively, the solution for $\Phi_*(t)$  is
\begin{equation}
\Phi_*(t)=\Phi_{\infty} \pm \frac13 \sqrt{\frac{6\Gamma}{\kappa \Psi^2_0}} \ ln{\left\{ th \left[\frac32 H_{\infty}(t-t_{*})\right]\right\}}\,,
\label{HH9}
\end{equation}
where $\Phi_{\infty} \equiv \Phi_{*}(t\to\infty)$ is the constant of integration. Using (\ref{HH6}) and (\ref{HH9}) we can reconstruct the function $\Phi_*(H)$ as follows:
\begin{equation}
	\Phi_*(H)= \Phi_{\infty} \mp \frac13 \sqrt{\frac{6\Gamma}{\kappa \Psi^2_0}} \ ln{\left[ \frac{\sqrt{H^2{-}H^2_{\infty}}{+}H}{H_{\infty}}\right]}\,.
\label{HH20}
\end{equation}
If we put $H=\sqrt{\frac13 \Omega^2}$, the reconstruction of the function $\Phi_*(\Omega^2)$ will be finished.

\subsubsection{Remark concerning the stage of fast Universe expansion}

When $t\geq t_0 > t_{*}$, the function $H(t)$, given by (\ref{HH6}), is monotonic with $\dot{H}<0$. It is interesting to compare the initial value $H(t_0)$ with $H(\infty)$. The corresponding ratio is
\begin{equation}
\frac{H(t_0)}{H(\infty)} = \sqrt{1+ \frac{\kappa \Psi^2_0}{2 \Lambda} {\dot{\Phi}}^2_{*}(t_0)}\,.
\label{0infty}
\end{equation}
If we assume that $\left|\dot{\Phi}_{*}(t_0)\right| >> \sqrt{\frac{2 \Lambda}{\kappa \Psi^2_0}}$, we obtain that the expansion of Early Universe is much faster than the late-time expansion.
This means that one can try to explain the inflationary stage using an appropriate fitting of the initial parameter $\dot{\Phi}_{*}(t_0)$.

\subsubsection{How to determine the auxiliary parameter $t_*$}

We assume, that Universe evolution starts at the time moment $t=t_0$ (in principle, one can consider $t_0=0$), and require that $t_0>t_{*}$. Thus, we avoid the model singularity, since for $t \geq t_0 > t_{*}$ the Hubble function takes only finite values. In other words, we consider the auxiliary parameter $t_{*}$ as a fictitious time moment, which does not belong to the time interval of the Universe evolution.

In order to identify the parameter $t_{*}<t_0$ and to choose properly the signs in (\ref{HH21}), we put $t=t_0$ into $\dot{\Phi}_{*}(t)$ given by (\ref{HH21}), and obtain the following three results.

\vspace{2mm}
\noindent
{\it (i) $\dot{\Phi}_*(t_0)$ is positive.}

\noindent
We take the sign plus in (\ref{HH21}) and obtain
$$
3H_{\infty}(t_0-t_*) =
$$
\begin{equation}
= ln \left[\frac{1}{\dot{\Phi}_{*}(t_0)} \sqrt{\frac{2\Lambda}{\kappa \Psi^2_0}} \left(1{+}\sqrt{1{+} \frac{\dot{\Phi}_{*}^2(t_0) \kappa \Psi^2_0}{2\Lambda}} \right) \right].
\label{HH12}
\end{equation}

\vspace{2mm}
\noindent
{\it (ii) $\dot{\Phi}_*(t_0)$ is negative.}
\noindent
We take the sign minus in (\ref{HH21}) and obtain (\ref{HH12}) modified by replacement $\dot{\Phi}_{*}(t_0) \to \left|\dot{\Phi}_{*}(t_0) \right|$; for sure, again we see that $t_0>t_{*}$.

\vspace{2mm}
\noindent
{\it (iii) $\dot{\Phi}_*(t_0)=0$.}
\noindent
This case can be realized, when $H_{\infty}=\infty$, i.e., $\Gamma {=} 0$, and thus, when $C_1{+}3C_2{+}C_3=-2$. We think this case to be non-physical.

\subsubsection{Remark about the transition point}

The acceleration parameter (\ref{HH8}) takes zero value at $t = t_{*} {+} \frac{1}{3H_{\infty}}\ ln [\sqrt3 \pm \sqrt2]$. Only one value of the cosmological time corresponds to inequality $t>t_{*}$, and we indicate the cosmological event at the time moment $t_{T}$:
\begin{equation}
t_{T} \equiv  t_{*} + \frac{1}{3H_{\infty}}\ ln [\sqrt3 + \sqrt2]
\label{HH8k}
\end{equation}
as the transition point, which separates the epochs of decelerated expansion $t_0<t<t_{T}$ and of accelerated expansion $t>t_{T}$.
For sure, we have to check, when $t_{T}>t_0$; this inequality holds, when $\left|\dot{\Phi}_*(t_0) \right|> \sqrt{\frac{\Lambda}{\kappa \Psi^2_0}}$.

\subsubsection{Searching for the components of the stress-energy tensor of the axionic field}

Since the axion potential vanishes on the solution $\phi = \Phi_{*}$, i.e., $V(\phi^2,\Omega^2)\equiv 0$, the axion energy density and pressure (\ref{TA1}) coincide
\begin{equation}
W_{(\rm A)} =  {\cal P}_{(\rm A)} = \frac{1}{2} \Psi^2_0 {\dot{\Phi}_{*}}^2 = \frac{\Lambda}{\kappa} sh^{-2}[3H_{\infty}(t-t_{*})] \,.
\label{TA11}
\end{equation}
This means that the effective equation of state for the axions is of the stiff type.

\subsubsection{Searching for the components of the stress-energy tensor of the vector field}

The aether energy density and pressure have the following form:
$$
W_{(\rm U)} = 3H^2(t)(1{-}\Gamma) \,,
$$
\begin{equation}
{\cal P}_{(\rm U)} = 3(1{-}\Gamma)[H^2(t)-2H^2_{\infty}] \,.
\label{TA29}
\end{equation}
The effective aether enthalpy
$$
{\cal P}_{(\rm U)} {+} W_{(\rm U)} = 6(1{-}\Gamma)(H^2{-}H^2_{\infty}) =
$$
\begin{equation}
=\frac{2\Lambda (1{-}\Gamma)}{\Gamma sh^2[3H_{\infty}(t{-}t_{*})]}
\label{TA27}
\end{equation}
is also proportional to the multiplier $1-\Gamma \equiv - \frac12 \left(C_1+ 3 C_2+C_3 \right)$. We have to discuss the sign of this cardinal parameter in the light of data obtained after the events GW170817 and GRB 170817A \cite{170817}.

\vspace{2mm}
\noindent
{\it (j) Are the gravitational wave modes in the aether subluminal?}

\noindent
If we interpret the results of observations \cite{170817}, assuming that
$1-3 \times 10^{-15}< \frac{v_{\rm gw}}{c}<1$, i.e., the velocity of gravitational wave mode in the aether is less than speed of light, we have to take into account the estimations of the gravitational Cherenkov effect (see, e.g., \cite{CH1,CH2}). Thus, according to \cite{constr1,constr2}, we have to consider $S_{(1)}>1$ and $S_{(0)}>1$ to avoid anomalous acceleration of cosmic particles, and obtain the following constraint for the Jacobson's parameter $C_2$: $0<C_2< 0.095$ \cite{constr1}. Taking into account that $|C_1+C_2|< 10^{-15}$, we see that in this case the parameter $1-\Gamma \simeq - \frac32 C_2$ is negative. This case relates to the negative energy density of the aether (see (\ref{TA29})).

\vspace{2mm}
\noindent
{\it (jj) Are the gravitational wave modes in the aether superluminal?}

\noindent

If we assume that $1 < \frac{v_{\rm gw}}{c}<1+ 7 \times 10^{-16}$, the argument concerning the gravitational Cherenkov effect can not be used, and the requirements that the phase velocities of the spin-1 and spin-0 modes are bigger than 1, are not necessary. In this case the constraint for $C_2$ is $-\frac{2}{27}<C_2<\frac{2}{21}$ (see \cite{constr1}), and we obtain the following inequality for the parameter $1-\Gamma$: $ - \frac17 <1-\Gamma < \frac19$. Clearly, there exists the possibility to have positive parameter  $1-\Gamma$ and positive aether energy density (\ref{TA29}).
When $\Gamma<1$, the aether pressure (\ref{TA29}) changes the sign at the moment $t_{d}$ defined as follows:
\begin{equation}
H(t_d)=\sqrt2 H_{\infty} \ \rightarrow t_d = t_{*} {+} \frac{1}{3H_{\infty}}  ln \left(\sqrt2{+}1\right) \,.
\label{TA28}
\end{equation}
At the same time moment $t{=}t_d$ the effective parameter $\zeta$, given by
\begin{equation}
\zeta(t) \equiv \frac{{\cal P}_{(\rm U)}}{W_{(\rm U)}} = 1-2 th^2[3H_{\infty}(t-t_{*})]\,,
\label{TA129}
\end{equation}
takes zero value, then changes the sign and tends asymptotically to minus one at $t \to \infty$.
One can say, that in the case  $\Gamma<1$ the aether asymptotically can play the role of the dark energy of the $\Lambda$ type.

\vspace{2mm}
\noindent
{\it (jjj) The case $\Gamma =1$.}

\noindent
There is a third very interesting situation, when $C_2=0$, $C_1+C_3=0$, so that  $\Gamma =1$. This is the case, when the velocity of the spin-0 aether mode is equal to zero, and the velocity of the spin-2 aether mode coincide with speed of light in vacuum. This model was studied in \cite{EM2}, where it was shown that in the framework of the model with strong gravity field the pp-wave aether modes can propagate if and only if $C_2=0$ and $C_1=-C_3$ simultaneously. Now we obtain that $W_{(\rm U)}=P_{(\rm U)}=0$, so that the aether is "invisible" from the energetic point of view.

\subsubsection{Asymptotic properties of the model functions}

The obtained solution has a quasi-de Sitter asymptote
$$
H(t \to \infty) \to H_{\infty} \,,  \quad -q(t \to \infty) \to 1 \,.
$$
\begin{equation}
 a(t \to \infty) \propto e^{H_{\infty} t} \,.
\label{HH71}
\end{equation}
The value $-q(t_0)$ is equal to
\begin{equation}
 -q(t_0) =  \frac{3H^2_{\infty}}{H^2(t_0)} -2 \,.
\label{HH75}
\end{equation}
Clearly, the transition point with $-q(t_T)=0$ exists, if $-q(t_0)$ is negative, while $-q(\infty)=1$ is positive. This is possible, when  $H(t_0)>\sqrt{\frac32} H_{\infty}$.
In the asymptotic limit $t \to \infty$ the axion energy density and pressure behave as follows:
\begin{equation}
W_{(\rm A)} =  {\cal P}_{(\rm A)} \propto e^{-6H_{\infty}t} \to 0 \,.
\label{HH73}
\end{equation}
In order to complete the analysis let us consider the case with vanishing cosmological constant.

\subsection{Exact solution, describing the basic state of axion field in the case $\Lambda = 0$}

When the cosmological constant vanishes, the solutions to the key equations for the Hubble function, scale factor and acceleration parameter take the form
\begin{equation}
H(t) = \frac{H(t_0)}{1+3H(t_0)(t-t_0)} \,, \quad H(t_0) = \sqrt{\frac{\kappa \Psi^2_0}{6\Gamma}} \  \dot{\Phi}_{*}(t_0)   \,,
\label{hh1}
\end{equation}
\begin{equation}
a(t) = a(t_0) \left[1+3H(t_0)(t-t_0)\right]^{\frac13} \,, \quad -q(t) = -2   \,.
\label{hh2}
\end{equation}
The basic state for the axion field is described by the function
\begin{equation}
\Phi_{*}(t) = \Phi_{*}(t_0) + \frac{\dot{\Phi}_{*}(t_0)}{3K}ln \left[1+ 3H(t_0)(t-t_0) \right]  \,.
\label{hh3}
\end{equation}
Reconstruction of the function $\Phi_{*}(H)$ yields
\begin{equation}
\Phi_{*}(H) = \Phi_{*}(t_0) - \frac{\dot{\Phi}_{*}(t_0)}{3H(t_0)}ln \frac{H}{H(t_0)}  \,.
\label{hh4}
\end{equation}
Now the axion energy density and pressure, and the corresponding aether quantities  are of the form
\begin{equation}
\kappa W_{(\rm A)} {=} \kappa {\cal P}_{(\rm A)} = \Gamma {\cal F} \,, \quad W_{(\rm U)} {=} {\cal P}_{(\rm U)} = (1{-}\Gamma) {\cal F} \,,
\label{TA11b}
\end{equation}
where
\begin{equation}
{\cal F} \equiv \frac{3 H^2(t_0)}{\left[1{+} 3H(t_0)(t{-}t_0) \right]^{2}} = \kappa W_{(\rm A)}{+} W_{(\rm U)}\,.
\label{TA411}
\end{equation}
Again, the aether is invisible, when $\Gamma=1$.

\subsection{Stability of the model with $\Lambda \neq 0$}

In order to solve the stability problem in a general form we have to go beyond the model in which all the quantities depend on time only, and thus we have to consider perturbations in an inhomogeneous universe. This problem is out of frame of this paper. However, one can answer the question: whether the homogeneous perturbations, depending on time only, can grow with cosmological time? For this purpose we assume that there are small variations of the pseudoscalar field and of the Hubble function:
\begin{equation}
\phi(t) \to \Phi_{*}(t) + \psi(t) \,, \quad H(t) \to H(t) + h(t)\,.
\label{pp1}
\end{equation}
Then the equations (\ref{App186}) and (\ref{App125}) give, respectively:
\begin{equation}
\frac{6\Gamma}{\kappa \Psi^2_0} H h = \dot{\Phi}_{*} \dot{\psi} + 2\gamma \psi \Phi^2_{*}H \frac{\ d \Phi_{*}}{dH}  \,.
\label{pp2}
\end{equation}
\begin{equation}
\ddot{\psi} + 3 H \dot{\psi} + 3h \dot{\Phi}_{*} + 2 \gamma \psi \Phi^2_{*} = 0  \,.
\label{pp3}
\end{equation}
When we extract $h$ from (\ref{pp2}), put it into (\ref{pp3}) and use (\ref{HH6}), (\ref{HH21}) and (\ref{HH20}) for $H$, $\dot{\Phi}_{*}$ and $\Phi_{*}$, we obtain the key equation
\begin{equation}
\ddot{\psi} + 3 H_{\infty}\dot{\psi} \left[ \frac{2H}{H_{\infty}} - \frac{H_{\infty}}{H} \right] = 0  \,.
\label{pp4}
\end{equation}
Surprisingly, this equation does not include $\psi$; first integration of (\ref{pp4}) yields
\begin{equation}
\frac{\dot{\psi}(t)}{\dot{\psi}(t_0)} = \left[\frac{sh[3H_{\infty}(t_0{-}t_{*})]}{sh[3H_{\infty}(t{-}t_{*})]} \right]^2
\left[\frac{ch[3H_{\infty}(t{-}t_{*})]}{ch[3H_{\infty}(t_0{-}t_{*})]} \right]  \,.
\label{3pp4}
\end{equation}
The ratio of the first derivatives of the perturbation and of the basic function
\begin{equation}
\frac{\dot{\psi}(t)}{\dot{\Phi}_{*}(t)} = \frac{\dot{\psi}(t_0)}{\dot{\Phi}_{*}(t_0)} \sqrt{\frac{2\Lambda+ \kappa \Psi^2_0 {\dot{\Phi}}^2_{*}(t)}{2\Lambda+\kappa \Psi^2_0 {\dot{\Phi}}^2_{*}(t_0)}}
\label{4pp4}
\end{equation}
does not grow with time; similarly to the function $\dot{\Phi}_{*}(t)$ (see (\ref{HH21})) the function $\dot{\psi}(t)$ vanishes asymptotically. The second integration gives
\begin{equation}
\psi(t) - \psi(t_0) = \frac{\dot{\psi}(t_0)}{3 H(t_0)}  \left[1 {-}\frac{\dot{\Phi}_{*}(t)}{\dot{\Phi}_{*}(t_0)} \right] \,.
\label{pp5}
\end{equation}
Clearly, the function $\psi(t)$ tends monotonically to the constant at $t \to \infty$:
\begin{equation}
\psi(\infty) = \psi(t_0) {+}
\frac{\dot{\psi}(t_0)}{3 H(t_0)} \,, \quad  |\psi(t){-} \psi(\infty)| \propto e^{{-}3H_{\infty}t}\,.
\label{pp5y}
\end{equation}
Linear approximation of the perturbation theory is valid, if not only $|\psi(t_0)| << |\Phi_{*}(t_0)|$, but also if $|\psi(\infty)| << |\Phi_{*}(\infty)|$; special case is when $\Phi_{*}(\infty)=0$, we have to choose $\psi(\infty)=0$ in this case. We see that there are no growing modes in the perturbations of pseudoscalar field. Moreover,
keeping in mind that
$$
|\dot {\Phi}_{*}(t\to\infty)| \propto e^{-3H_{\infty}t} \,, \quad \Phi^2_{*} \propto e^{-6H_{\infty}t} \,,
$$
\begin{equation}
\left|H \frac{\ d \Phi_{*}}{dH}\right|(t\to\infty) \propto e^{3H_{\infty}t}\,,
\label{pp5b}
\end{equation}
we find that the asymptotic behavior of the function $h(t)$ is $|h(t\to\infty)| \propto e^{-3H_{\infty}t}$, when $\psi(\infty)\neq 0$, and
$|h(t\to\infty)| \propto e^{-6H_{\infty}t}$, when $\psi(\infty) = 0$.
The corresponding estimations for the scale factor are, respectively
$$
|a(t\to\infty) -a_{\infty}e^{H_{\infty}t}| \propto e^{-2H_{\infty}t} \ \ (\psi(\infty)\neq 0)\,,
$$
\begin{equation}
 |a(t\to\infty) -a_{\infty}e^{H_{\infty}t}| \propto e^{-5H_{\infty}t} \ \ \ (\psi(\infty) = 0)\,.
\label{pp5m}
\end{equation}
We deal with asymptotically stable model.

\section{Discussion}

When we prepared the manuscript, we kept in mind two models of stiff control in moving physical systems. The first model deals with a moving dense plasma in external fields (gravitational and/or electromagnetic); in this model frequent particle collisions enforce the distribution function of plasma to follow the specific equilibrium function, which turns to zero the collision integral \cite{Groot}, and depends on parameters of external fields. The second model relates to the dynamics of large granules in the viscous fluid flow; the Stokes force in this model coerces the granules to have the velocity coinciding with the non-uniform macroscopic velocity of the fluid flow \cite{Jou}. We tried to imagine, how the dynamic aether could realize the stiff regulation of the behavior of axionic dark matter. These two analogies hinted us, that the guidance of such kind is possible trough the specific Higgs potential, $V(\phi^2,\Omega^2 )= \frac12 \gamma \left[\phi^2 {-} \Phi^2_{*}(\Omega^2) \right]^2$, describing non-linear self-interaction of pseudoscalar (axion) field. When the Higgs potential turns into zero, we obtain the analog of equilibrium for the axionic system in the aether flow. When the  basic state $\Phi_{*}(\Omega^2)$ is not constant and depends on the aether guiding function $\Omega^2 \equiv g^{mn}g^{pq}\nabla_m U_p \nabla_n U_q$, we face with the  aether control over the state of the dark matter. The appropriate tool for this task is the axionic extension of the Einstein-aether theory; the corresponding extension is connected with the fact that now the potential $V(\phi^2,\Omega^2)$ includes not only the pseudoscalar field in square $\phi^2$, but also the vector field, the aether velocity four-vector, the metric and Christoffel symbols. The variation procedure gives the corresponding additional source terms into the master equations of the vector field (\ref{0A1}), (\ref{J2}) and of the gravitational field (see, e.g., (\ref{5Ein1}) with (\ref{J29})).

Since the cosmology is the natural application of this model, we reduced the obtained master equations to the symmetry of homogeneous isotropic Friedmann - type model. In this model the guiding function $\Omega^2$ is proportional to the square of the Hubble function $\Omega^2=3H^2$, thus, just the rate of cosmological expansion predetermines the evolution of the basic state of the pseudoscalar field. We have found exact solutions of the reduced system of master equations in case of stiff regulation, i.e., when the value of the pseudoscalar field $\phi(t)$ at any time $t$ coincides with the basic state function $\Phi_{*}(H(t))$. When the cosmological constant is non-vanishing, $\Lambda \neq 0$, the Hubble function is presented by (\ref{HH6})), and the scale factor is given by (\ref{HH7}); when $\Lambda= 0$ we deal with the formulas (\ref{hh1})) and (\ref{hh2}), respectively. The quantity $\Phi_{*}$ as a function of cosmological time is given, respectively, by (\ref{HH9}) and (\ref{hh3}); reconstruction of the function $\Phi_{*}(H)$ gives, respectively, (\ref{HH20}) and (\ref{hh4}).

 Let us recall that introduction of the cosmological constant $\Lambda$ into the model does not guarantee, in general case, that the asymptotic behavior of the model is of the de Sitter type (see, e.g., \cite{Volterra} for the examples of such behavior). In addition, we have to emphasize that many cosmological models are characterized by the so-called quasi-de Sitter asymptotic regime, when the Hubble function tends to constant at $t \to \infty$, i.e., $H \to H_{\infty}$. In the de Sitter case the Hubble function is constant, $H(t)= H_{\infty} = \sqrt{\frac{\Lambda}{3}}$, and it is connected with the cosmological constant $\Lambda$. In general case
$H_{\infty} \neq \sqrt{\frac{\Lambda}{3}}$, and there is a very interesting and non-trivial question: how does the quantity $H_{\infty}$ depend on the physical parameters of the corresponding model (for instance, on the coupling constant $\Psi_0$ or on the initial value $\dot{\Phi}_{*}(t_0)$, which appeared in the axionic extension of the Einstein-aether model).  In this sense, we consider the quasi-de Sitter asymptotic behavior as a quality indicator of the model, which explains the late-time accelerated expansion of the Universe, and avoids the catastrophic scenaria of the Universe final stage.

Finally, we have to formulate the following conclusions concerning the model in the framework of which the dynamic aether is shown to provide stiff regulation of the axionic dark matter behavior.

\vspace{1mm}
\noindent
1. The axionic extension of the Einstein-aether theory with cosmological constant $\Lambda \neq 0$ guarantees the existence of one transition point in the Universe history at $t=t_{T}$ (\ref{HH8k}), which separates the epochs of decelerated expansion and of the late-time accelerated expansion; as well as, it guarantees that the asymptotic regime of expansion is of the quasi-de Sitter type (Pseudo-Rip) with Hubble constant $H_{\infty}{=} \sqrt{\frac{\Lambda}{3 \Gamma}}$, where $\Gamma {=} 1{+}\frac12 (C_1{+}3C_2{+}C_3)$ is the parameter containing three Jacobson's coupling constants $C_1,C_2,C_3$. According to the data published after the events encoded as GW170817 and GRB 170817A \cite{170817}, $|C_1+C_3|<10^{-15}$, so the parameter $C_2$ is the crucial one for the determination of the parameter $\Gamma$. Strictly speaking, estimations admit all three possibilities: $\Gamma <1$, $\Gamma =1$ and $\Gamma >1$. When $\Gamma =1$, the aether behaves as an invisible substratum, since the aether energy density and pressure are equal to zero. The case $\Gamma <1$ is also possible and can be considered as physically motivated, since the energy density of the aether is positive.

\vspace{1mm}
\noindent
2. When $\Gamma <1$, the scalars, which describe the effective energy density and pressure of the aether (\ref{TA29}), demonstrate four interesting features: {\it first}, the aether energy density $W_{(\rm U)}$ is positive, when the effective Jacobson's parameter $C_1{+}3C_2{+}C_3$ is negative and satisfies the inequality $-2<C_1{+}3C_2{+}C_3<0$; {\it second}, the effective aether pressure ${\cal P}_{(\rm U)}$ changes the sign and becomes negative at the time moment $t=t_d$ (see (\ref{TA28})); {\it third}, the function $W_{(\rm U)}{+}3{\cal P}_{(\rm U)}$ becomes negative, when $t>t_T>t_d$; {\it fourth},  the effective aether enthalpy $W_{(\rm U)}{+}{\cal P}_{(\rm U)}$ tends to zero asymptotically at $t\to \infty$ (see (\ref{TA27})). In other words, the aether behaves as a {\it dark energy}, starting from the time moment $t_T$, and asymptotically becomes the dark energy of the $\Lambda$ type. When $\Lambda {=} 0$, the model can not explain the late-time accelerated expansion, since the acceleration parameter is negative for any time moment.

\vspace{1mm}
\noindent
3. The scalars of energy density and pressure of the axionic dark matter (\ref{TA11}) demonstrate that the corresponding effective equation of state, $W_{(\rm A)} {=} {\cal P}_{(\rm A)} >0$, can be indicated as the stiff one; this fact attracts the attention to the studies of the so-called {\it stiff eras} appeared in the framework of Modified Gravity (see, e.g., \cite{stiff} and references therein).

\vspace{1mm}
\noindent
4. When $\Lambda \neq 0$, homogeneous perturbations of the pseudoscalar (axion) field, of the Hubble function and of the scale factor fade out with cosmological time,
i.e., we deal with the {\it stable} model of stiff regulation of the axionic dark matter behavior by the dynamic aether.

\vspace{5mm}

\acknowledgments{The work was supported by Russian Science Foundation (Project No. 16-12-10401), and, partially, by the Program of Competitive Growth
of Kazan Federal University.}

\end{document}